\documentclass[a4paper,12pt]{article}
\usepackage{amssymb}
\usepackage{graphicx}
\usepackage{booktabs}
\usepackage{array}
\usepackage{multirow}
\usepackage{hyperref}
\usepackage{cite}
\usepackage{appendix}
\usepackage{amsmath}
\usepackage{geometry}
\usepackage{enumitem}
\usepackage{float}
\usepackage{caption}
\usepackage{graphicx}
\usepackage{tabularx}
\usepackage{makecell}
\begin{document}
\title{\bf{Numerical evaluation of parton distribution Mellin moments}}

\author{ $ \mathrm{\textbf{Akbari Jahan}^{1\star}} $ and $\mathrm{\textbf{Diptimonta Neog}^{2}}$ \\ \small{Department of Physics, North Eastern Regional Institute of Science and Technology}\\ \small{Nirjuli-791109, Arunachal Pradesh, India.}\\ \small { $ {}^{\star} $ Email: akbari.jahan@gmail.com}}
\date{}

\maketitle

\begin{abstract}
The detailed comprehension of momentum fraction and energy dependence of proton structure functions is among the major difficulties in high-energy physics. Perturbative quantum chromodynamics (QCD) plays as an extensive foundation for analysing deep inelastic scattering (DIS) and experiments with hadron-hadron collisions. In such a framework, where the hard collisions advance through the partonic components of hadrons, a pivotal contribution is imparted by parton distribution functions (PDFs). PDFs provide information about the nucleonic substructure in terms of its components.  PDFs have become more and more convictive following the advent of the Large Hadron Collider (LHC). Moments of PDFs are important quantities for the analysis of hadronic internal structure. The Mellin moments of such parton densities have been discussed and their analyses have been carried out numerically.\\



\end{abstract}


\section{Introduction}

Parton Distribution Functions (PDFs) are functions that elucidate the probability of momentum fraction of proton carried by the constituent quarks and gluons. The collision of protons at high energies, as in the Large Hadron Collider (LHC), provides an important information about their energy and momentum contents. However, the exact calculation of PDFs are indeterminable using perturbative quantum chromodynamics (QCD), which is accountable for holding the partons together. This is due to the fact that partons are inherently non-perturbative in nature and the computation of such non-perturbative PDFs is difficult at the proton scale. Several physical processes are involved in the determination of PDFs which is a complex effort. Deep inelastic scattering (DIS) experiments as well as experiments run at the DESY (Deutsches Elektronen Synchrotron) laboratory were basically established to measure the PDFs of proton. Without PDFs, theoretical conjecture at a hadron collider is not possible \cite{1, 2}. However, parton distributions cannot be calculated using the first principles with our existing know-how of strong interactions. They can, nevertheless, be determined only from data sets using suitable theoretical presumptions. The effect of such presumptions is fairly minimized by using maximum number of data \cite{3,4,5,6}. Such PDFs, dependent on proton energy scale, obtained at one energy scale can be measured at another energy scale by using DGLAP equations \cite{7,8,9,10}. These equations thus evolve PDFs from some starting energy scale, say $Q_0$, to any higher scale. A number of collaborations across the globe are assigned for determining PDFs. NNPDF collaboration, which uses neural networks to extricate the PDFs and Monte Carlo methods to determine their statistical distribution, is one of the largest and the most successful of all. Currently, PDF sets in NNPDF, MSHT and CT collaborations, with all the required features, are being maintained and updated.\\

A parton distribution function $f_i(x,Q)$, where \textit{i} represents the parton species, depends on two variables: parton momentum fraction \textit{x} possessed by the parton and momentum scale \textit{Q} at which the nucleon is observed. Since PDFs are essential for the explication of present and future collider experiments, good theoretical predictions of uncertainty analysis can be developed using the greater accurate data available. Gluons are responsible for carrying approximately half of the proton momentum. While up and down quarks dominate at large $ x $, gluons dominate and can even range negative at small $ x $. Gluon distribution increases with the increase in energy scale $ Q $. Partonic interactions result in some significant elementary processes such as gluon formation from quarks and gluons ($q \rightarrow qg$ and $ g \rightarrow gg $) as well as quark-antiquark pair formation by gluons ($ g \rightarrow q\bar q $) \cite{11}. These cause parton distributions in QCD scale dependent and is valid in all orders in the running coupling constant ($ \alpha_s $). It is steered by a set of coupled integro-differential QCD evolution equations \cite{7,8,9,10}. It also has the basic characteristic that $ f_i(x,Q) $ is a declining function of $Q$ at large $x$, while at small $x$, it becomes a growing function \cite{12, 13}.\\

For asymptotically large $x$, PDFs perform as $(1-x)^{b_i}$, $b_i$ being derived from the Brodsky-Farrar quark counting rules \cite{14}. On the other hand, proton PDFs act as $x^{a_i}$ at asymptotically small $x$, $a_i$ being deduced from the Regge theory \cite{15}. Several theoretical constraints are used to determine PDFs \cite{16}. Conservation of energy suggests that the momentum fraction $x$ possessed by each unpolarised parton must add up to one. PDFs satisfy several momentum and number sum rules, and thus provide useful cross check on the results. The Bjorken sum rule \cite{17} cites to the first moment of the spin-dependent nonsinglet structure function. Sum rules are effective mechanisms to examine the nucleonic internal spin structure. One of such momentum sum rules is as follows: 

\begin{equation}
\label{sum rule}
\sum_i \int_0^1 x \, f_i(x, Q) dx=1
\end{equation}

Furthermore, valence sum rules confirm the valence structure of an unpolarized hadron; that is, if we consider a proton that is comprised of one down quark and two up quarks, we can have:
\begin{eqnarray}
\label{array}
\int_0^1 (u-\bar u) dx & = & 2 \nonumber \\
\int_0^1 (d-\bar d) dx & = & 1\nonumber \\
\int_0^1 (s-\bar s) dx & = & 0 
\end{eqnarray}

In other words, conservation of baryon number in proton gives

\begin{equation}
\label{baryon number}
\int_0^1 [u(x,Q^2)- \bar u (x,Q^2)] dx = 2 \int_0^1 [d(x,Q^2)- \bar d (x,Q^2)] dx = 2
\end{equation}

and conservation of total energy-momentum is given as

\begin{equation}
\label{energy-momentum}
\int_0^1 x \left[ \sum_{i=1}^{n_f} \left(q_i(x,Q^2)+ \bar q_i (x,Q^2) \right) + g(x,Q^2) \right] dx=1
\end{equation}

where $n_f \equiv $ number of quark flavours.\\

Evidently, these sum rules provide limitations on the behaviour of parton distributions even in the domain where data are not available \cite{18}. Parton distributions can be ascertained by determining at least seven independent functions: three quark and three antiquark distributions as well as the gluon at some initial scale, from which PDFs at all other scales can be obtained using evolution equations. PDFs are, therefore, in general, assumed to have the form
\begin{equation}
\label{pdf}
f_i(x,Q^2)=x^{a_i} (1-x)^{b_i} C(x)
\end{equation}
where at $x \rightarrow 0$ and $ x \rightarrow 1$, $C(x) \rightarrow$ constant.\\

The behaviour of partons at both large $x$ and small $x$ are relatively connected by momentum sum rule. The established DGLAP equations govern the parton distributions and their moments, viz. sum rules and spin of proton. It is attained by integrating the parton distributions over momentum fraction $x$. Alternately, the evolution of Mellin moments of the parton distributions can be directly studied \cite{19}. The changes in parton distributions arise due to the hard interactions between quarks and gluons. As mentioned before, the evolution of PDFs is characterised by the basic DGLAP equations:

\begin{equation}
\label{DGLAP}
\frac{dq(x,Q^2)}{d \ln Q^2}=\frac{\alpha_s(Q^2)}{2\pi}(P\otimes q)(x,Q^2) 
\end{equation}

where $\alpha_s(Q^2)$ is the running coupling constant, $P(t)$ is the splitting function and $\otimes$ represents the Mellin convolution:

\begin{equation}
(P \otimes q)(x) \equiv \int_x^1 \frac{dt}{t} P(x/t) \, q(t) 
\end{equation}

Several methods are available in literature to obtain the solution of the integro-differential DGLAP equations, which are based on Mellin transformation technique, method of expansion of polynomials, etc. In our present work, the Mellin transformation method \cite{20} has been adopted to obtain the numerical evaluation and analysis in moment space. The integro-differential equations become very simple using this transformation method.

\section{Mathematical Framework}
\textbullet \quad \textbf{Definition of Mellin moments} \\

The qualitative probe of PDFs involves the application of Mellin moments \cite{21, 22}, which is instrumental in understanding the proton structure function. While PDF Mellin moments have been explored analytically in Ref. \cite{23}, Mellin transforms of some integral functions have been examined in Ref. \cite{24}. The $n$th moment evaluation of self-similarity based PDFs was also explored in Ref. \cite{25}. Using Eq. \ref{pdf}, the $n$th Mellin moment $f_n$ of a distribution function $f(x)$ may be determined as \cite{26}

\begin{equation}
\label{Mellin}
M f_n= \int^1_{0}  x^{n-1} \, f(x) \, dx
\end{equation}

where $n$ is the moment index. Now, for a given integral of the form in Eq. \ref{Mellin}, the following property of Gamma function may be used for a parton specie $i$ as follows:

\begin{equation}
\label{Gamma}
\int^1_0 x^{a_i} (1-x)^{b_i} \, dx= \frac {\Gamma(a_i+1) \, \Gamma(b_i+1)}{\Gamma(a_i+b_i+2)}
\end{equation}
\\
Regge theory suggests that $x\, f_i \sim (1-x)^{2n_s -1}$, that is, $b_i=2n_s-1$, where $n_s$ is the number of spectator quarks. The central fit has $\alpha_s(M_z)=0.118$ \cite{27}.\\\\

\textbullet \quad \textbf{Mellin moments of parton distributions} \\

The Mellin moments of parton distributions can thus be obtained by subsequently substituting the values of moment index $n$ in Eqs.  \ref{Mellin} and  \ref{Gamma} as follows:

\begin{equation}
\mathrm{For} \,\,\,  n=1: \quad Mf_1=\int_0^1  x^0 \, f(x) \, dx = \int_0^1 x^{a_i} \, (1-x)^{b_i} \, dx = \frac{\Gamma (a_i+1) \,  \Gamma (b_i+1)}{\Gamma (a_i+b_i+2)}  
\end{equation}

\begin{equation}
\mathrm{For} \,\,\,  n=2: \quad Mf_2=\int_0^1  x^1 \, f(x) \, dx = \int_0^1 x^{a_i+1} \, (1-x)^{b_i} \, dx = \frac{\Gamma (a_i+2) \, \Gamma (b_i+1)}{\Gamma (a_i+b_i+3)} 
\end{equation}

\begin{equation}
\mathrm{For} \,\,\, n=3: \quad Mf_3=\int_0^1  x^2 \, f(x) \, dx = \int_0^1 x^{a_i+2} \, (1-x)^{b_i} \, dx = \frac{\Gamma (a_i+3) \, \Gamma (b_i+1)}{\Gamma (a_i+b_i+4)} 
\end{equation}

\begin{equation}
\mathrm{For} \,\,\, n=4: \quad Mf_4=\int_0^1  x^3 \, f(x) \, dx = \int_0^1 x^{a_i+3} \, (1-x)^{b_i} \, dx = \frac{\Gamma (a_i+4) \, \Gamma (b_i+1)}{\Gamma (a_i+b_i+5)} 
\end{equation}

\begin{equation}
\mathrm{For} \,\,\, n=5: \quad Mf_5=\int_0^1  x^4 \, f(x) \, dx = \int_0^1 x^{a_i+4} \, (1-x)^{b_i} \, dx = \frac{\Gamma (a_i+5) \,  \Gamma (b_i+1)}{\Gamma (a_i+b_i+6)} 
\end{equation}

and so on.\\

With the resummation of double logarithms \cite{28}, the following values of $a_i$ and $b_i$ have been identified.
\begin{enumerate}[label=(\roman*)]
\item For gluon distribution: $a_i=-0.25 \, ,  \, b_i=3$.
\item For sea quarks: $a_i=-0.2 \, ,  \, b_i=4$.
\item For valence quarks: $a_i=0.63 \, ,  \, b_i=1$.
\end{enumerate}

Mellin moments of PDFs for a few values of $n$ have been listed in Table \ref{Table1}. As the moment index $n$ increases, Mellin moments  are found to decrease gradually.

\begin{table}[H]
\caption{Mellin moments of gluons, sea quarks and valence quarks for some values of $n$.}
\label{Table1}
\centering
\begin{tabular}{|c|c|c|c|}
\hline
\multirow{2}{*}{\textbf{Moment index ($n$)}} &  \multicolumn{3}{c|}{$M f_n$}\\
\cline{2-4}
&  \textbf{For gluon distribution} & \textbf{For sea quarks} & \textbf{For valence quarks} \\
\hline
1 & 0.443139 & 0.326337 & 0.233267 \\
2 & 0.069993 & 0.045012 & 0.104746 \\
3 & 0.021302 & 0.011915 & 0.059499 \\
4 & 0.008679 & 0.004277 & 0.038363 \\
5 & 0.004199 & 0.001847 & 0.026790 \\
\hline
\end{tabular}
\end{table}

On the other hand, using Ref. \cite{29}, we identify with a different set of values for $a_i$ and $b_i$. The sum of valence quarks in a proton is 3: two of them are up quarks and one of them is down quark. The net number of antiquarks is 0. The parameter $b_i=2n_s-1$, with $n_s$ being the minimum number of spectator quarks, is estimated from quark counting rules. Therefore, for valence quarks in a proton (\textit{qqq}), $n_s=2\, , \, b_i=3$; for gluon in a proton (\textit{qqqg}), $n_s=3 \, , \, b_i=5$ and for antiquarks in a proton $(qqqq\bar q), \, n_s=4 \, , \, b_i = 7$. The parameter $a_i$ is estimated from Regge arguments. Gluons and antiquarks have $a_i \sim -1$ while valence quarks have $a_i \sim 0.5$.

\begin{enumerate}[label=(\roman*)]
\item For gluon distribution: $a_i=-1 \, ,  \, b_i=5$.
\item For sea quarks: $a_i=-0.5 \, ,  \, b_i=3$.
\item For valence quarks: $a_i=-1 \, ,  \, b_i=7$.
\end{enumerate}

The values of Mellin moments for the above set of partons are listed in Table \ref{Table2}. In this case also, the values of Mellin moments decrease gradually as $n$ increases. However, for $n=1$, the values are not defined for gluon distribution as well as for antiquarks. \\

\begin{table}[H]
\caption{Mellin moments of gluons, sea quarks and antiquarks for some values of $n$.}
\label{Table2}
\centering
\begin{tabular}{|c|c|c|c|}
\hline
\multirow{2}{*}{\textbf{Moment index ($n$)}} &  \multicolumn{3}{c|}{$M f_n$}\\
\cline{2-4}
&  \textbf{For gluon distribution} & \textbf{For sea quarks} & \textbf{For antiquarks} \\
\hline
1 & $\infty$ & 0.101588 & $\infty$ \\ 
2 & 0.166667 & 0.027706 & 0.125 \\
3 & 0.023809 & 0.010656 & 0.013889 \\
4 & 0.005952 & 0.004973 & 0.002778 \\
5 & 0.001984 & 0.002633 & 0.000758 \\
\hline
\end{tabular}
\end{table}

\section{Results and Discussion}

\begin{figure}[h]
\centering
\includegraphics[scale=.28]{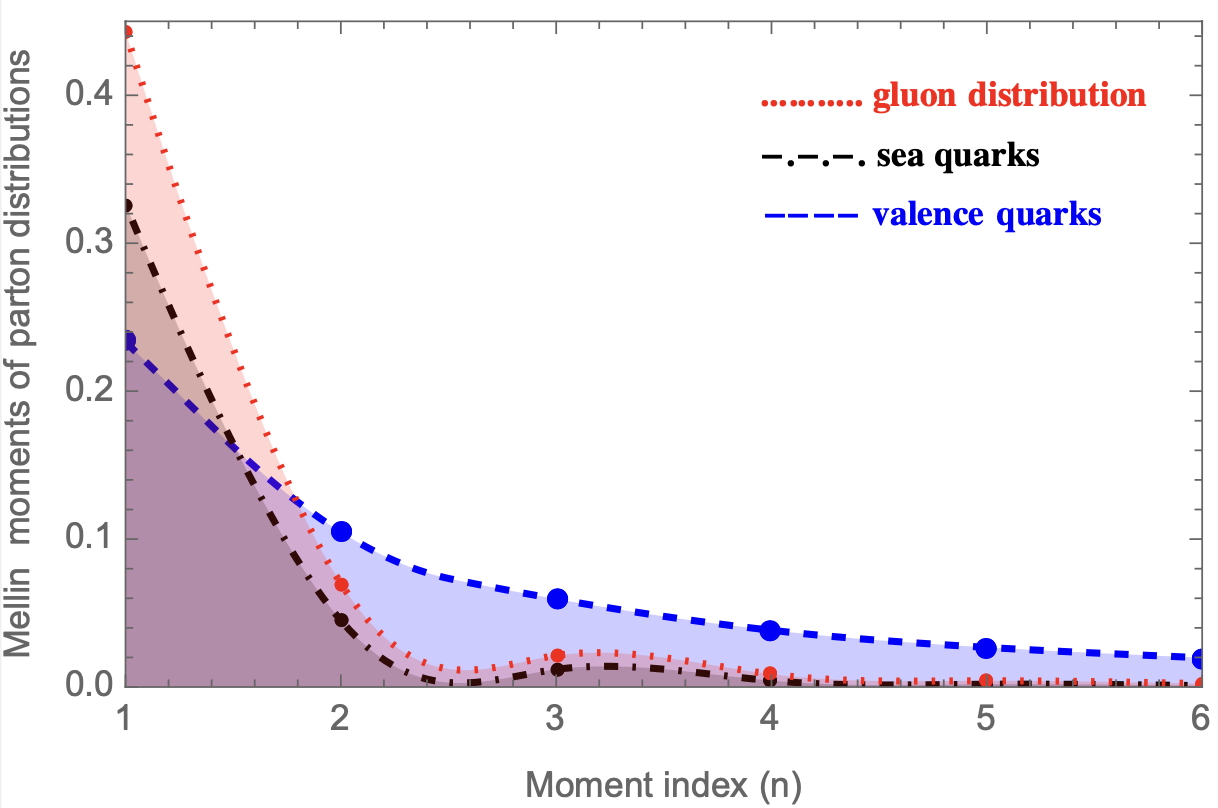}
\caption{Plot of Mellin moments of parton distributions against moment index.}
\label{fig:Mellin}
\end{figure}

When plotting the Mellin moments against the moment index $n$, as shown in Fig. \ref{fig:Mellin}, the horizontal axis represents the moment index $n$ and the vertical axis represents the values of the Mellin moments. The plot provides insight into the behaviour  of parton distributions. The moments generally decrease with increasing $n$ because higher moments emphasise on smaller $x$ values, where the distribution tends to be smaller. The Mellin moments of the gluon distribution indicate how gluons contribute to the hadron’s momentum. Since gluons dominate at small $x$, the Mellin moments for gluons typically start high for small $n$ and decrease rapidly as $n$ increases. Sea quarks are quark-antiquark pairs that are created and annihilated within the hadron. The curve for sea quarks drops more gradually, indicating their prevalence at smaller $x$ values. In other words, they contribute significantly at small $x$. Those quarks that determine the quantum numbers of hadrons are the valence quarks. Valence quark distributions typically peak at intermediate $x$ values and decrease at very large and small $x$. The curves for valence quarks show a slower decline, reflecting their significant presence at intermediate $x$ values.\\

The three different curves in Fig. \ref{fig:Mellin} represent different types of partons. By comparing Mellin moment curves for different types of partons, one can understand the corresponding contributions of gluons, sea quarks as well as valence quarks to the structure function of proton at various momentum fraction ranges.\\

Table \ref{Table3} shows the comparison of our method and key results on PDFs with those from other studies.

\begin{table}[H]
\caption{A comparative analysis of PDFs by different studies.}
\label{Table3}
\centering

\resizebox{\textwidth}{!}{
\begin{tabular}{|c|c|c|c|}
\hline
\textbf{Study} & \textbf{Method} & \textbf{Key results} & \textbf{Uncertainties} \\
\hline
\makecell{Forte \& Watt (2013) \\ (Ref. \cite{1})} & Global Analysis & Gluon distribution at $x \sim 0.01$ & $\pm 5\%$ \\
\hline
\makecell{Martin \textit{et al.} (1988) \\ (Ref.\cite{3})} & Structure function analysis & Valence quarks at large $x$ & $\pm 10\%$ \\
\hline
Current Work & Mellin Transformation & Gluon moments decrease rapidly & \makecell{$\pm 3\%$ \\ (from data fitting)}\\

\hline
\end{tabular}
}
\end{table}

Further comparison on PDF calculations with errors, obtained using different methods by different authors, has been done and listed in Table \ref{Table4}.

\begin{table}[H]
\caption{Comparison of PDF calculations obtained by different methods.}
\label{Table4}
\centering

\resizebox{\textwidth}{!}{
\begin{tabular}{|c|c|c|c|c|}
\hline
\textbf{Source} & \textbf{Data type} & \textbf{Method/Results description} & \textbf{Reference} & \textbf{Errors/Uncertainties} \\
\hline
\makecell{Current \\ work} & \makecell{Mellin Moments \\ of PDFs} & \makecell{Numerical evaluation of Mellin \\ moments for quarks and gluons} & Our work & \makecell{Data fitting and \\ Truncation error} \\ 
\hline
HERA & \makecell{DIS \\ cross sections} & \makecell{Measurements of structure \\ functions from DIS experiments} & Ref. \cite{30} & \makecell{Systematic \\ uncertainties} \\
\hline
NNPDF & \makecell{PDFs for LHC \\ Run II} & \makecell{Neural network fitting of PDFs; \\ comprehensive uncertainties} & Ref. \cite{35} & \makecell{Statistical and \\ systematic errors} \\
\hline
MMHT & \makecell{PDFs with \\ high precision} & \makecell{Global analysis incorporating \\ multiple data sources} & Ref. \cite{36} & \makecell{Detailed uncertainty \\ analysis} \\
\hline
CT14 & \makecell{Global PDF \\ analysis} & \makecell{Fixed-order QCD analysis \\ using a large dataset} & Ref. \cite{37} & \makecell{Comprehensive \\ error estimates} \\
\hline
ABM & \makecell{PDFs from a \\ global analysis} & PDFs tuned to LHC data & Ref. \cite{38} & Statistical uncertainties \\

\hline
\end{tabular}
}
\end{table}

Our work offers a comprehensive analysis across a wider range of momentum fractions $x$. By systematically varying the moment index $n$, we are able to provide insights into parton distributions at both small and large $x$, filling gaps left by previous studies. Unlike previous analyses that focused predominantly on large $x$ behavior, our calculations provide detailed insights into the gluon and sea quark contributions at small $x$, which are crucial for understanding high-energy scattering processes at the LHC. Our findings reveal that the Mellin moments for gluons decrease more rapidly than previously anticipated as $n$ increases, indicating a need for revised models in the small $x$ regime.

\section{Conclusions and outlook}

PDFs are a necessary framework for the study of precision Standard Model and pursuance of new physics at hadron colliders as well as at experiments using hadron targets. Understanding the  structure of proton is of immense significance, as protons are employed to attain the high-energy confines in hadron colliders such as the LHC \cite{30}. Determination of PDFs is a key to understanding LHC physics, without which there is no method of reliably exploring possible new physics \cite{1, 31,32,33,34}. Since PDFs are immeasurable in perturbation theory, they must be surmised from experimental data. It is paramount to accurately compute the PDFs of protons, whose unpredictability is a delimiting component in the precision of QCD theoretical conjecture. Owing to the present limitations in understanding the non-perturbative QCD, such an ascertainment is unachievable from the first principles. PDFs are, therefore, computed using perturbative QCD in global fits of hard-scattering experimental data \cite{35,36,37,38,39}. The standard approach to PDF evaluation is done by assuming for PDFs, at some reference energy scale $Q_0$, a functional form governed by counting rules, which suggest that PDFs behave as $f_i(x) \underset{x\to 1}{\sim} (1-x)^{b_i} $, and Regge theory, which suggest that they behave as $f_i(x) \underset{x\to 0} {\sim} x^{a_i}$ \cite{40}. Using Mellin transformation method and Gamma function property, computation of Mellin moments of such PDFs has been carried out. The Mellin moments thus computed show that their values drop gradually as the moment index $n$ increases.\\

While previous works have shown a significant contribution from sea quarks at intermediate $x$, our analysis reveals that gluons also play a more critical role at small $x$, which is essential for accurately modeling high-energy collisions. Recent studies utilizing lattice QCD \cite{41, 42} have provided complementary insights into the PDF structure, yet the non-perturbative nature of PDFs at small $x$ remains a challenging area, underscoring the importance of our approach in filling these gaps. The comprehensive numerical evaluation of Mellin moments presented in this work not only enhances the understanding of parton distributions but also provides a robust framework for future analyses that aim to refine predictions at hadron colliders. The present work aims to provide a clear and accessible method for evaluating Mellin moments of parton distributions, focusing on elucidating fundamental trends and behaviours. Simplified calculations can serve as valuable stepping stones, allowing for the establishment of a foundational understanding that can be built upon with more complex models in future work. Our findings can be seen as an invitation for further research, encouraging the integration of experimental data and more complex computational techniques. We believe that this iterative process is vital for refining our understanding of PDFs in the context of perturbative QCD.\\

Further work will be carried out taking into account the inclusion of a smooth polynomial rather than the constant $C(x)$ in Eq. \ref{pdf}. The results will then be compared with the present sets of values obtained in this paper. The broader pertinence of this method also needs to be examined more.

\section*{Acknowledgement}

The authors thank Prof. (Retd.) D. K. Choudhury, Gauhati University for his valuable suggestions during the preparation of the manuscript.

\section*{Declaration of competing interest}
The authors declare of not having any competing interests relevant to the subject matter of this manuscript. There has been no funding received for carrying out this work.

\section*{Data availability}
No data is to be deposited. The requisite information is provided in the particulars documented above.

\end{document}